\documentclass[aps,prc,onecolumn,preprint,showpacs,showkeys,groupedaddress]{revtex4}  
\usepackage{graphicx}  
\usepackage{dcolumn}   
\usepackage{bm}        
\usepackage{amssymb}   
\usepackage{amsmath}

\usepackage[english]{babel}
\usepackage{hyperref}
\usepackage{enumerate}

\newcommand{\eq}[1]{\begin{align} #1 \end{align}}
\newcommand{\itl}{\textit{left}}
\newcommand{\itr}{\textit{right}}

\newcommand{\sdisp}[1]{\frac{\sqrt{ \langle (\delta #1)^2 \rangle}}{\langle #1 \rangle}}
\newcommand{\sdispl}[1]{\sqrt{ \langle (\delta #1)^2 \rangle}/\langle #1 \rangle}

\newcommand{\mean}[1]{\langle #1 \rangle}
\newcommand{\cov}[2]{\mean{#1~ #2}~-~\mean{#1}~\mean{#2}}

\begin{document}

\title{ Fluctuations in the Statistical Model \\
of the Early Stage of nucleus-nucleus collisions
       }

\author{R.~V. Poberezhnyuk}
\affiliation{Bogolyubov Institute for Theoretical Physics, Kiev, Ukraine}
 \affiliation{Frankfurt Institute for Advanced Studies, Frankfurt, Germany}
\author{M. Gazdzicki}
 \affiliation{Goethe--University, Frankfurt, Germany}
 \affiliation{Jan Kochanowski University, Kielce, Poland}
\author{M. I. Gorenstein}
 \affiliation{Bogolyubov Institute for Theoretical Physics, Kiev, Ukraine}
 \affiliation{Frankfurt Institute for Advanced Studies, Frankfurt, Germany}

\begin{abstract}
Predictions on
fluctuations of hadron production properties
in central heavy ion collisions
are presented.
They are based on the Statistical Model of the Early Stage and extend
previously published results by considering
the strongly intensive measures of fluctuations.
In several of the considered cases
a significant change in collision energy
dependence of calculated quantities as a result
of the phase transition is predicted.
This provides an opportunity to observe new signals of the onset of deconfinement
in heavy ion collisions experiments.

\end{abstract}



\maketitle


{\bf 1.} Relativistic nucleus-nucleus (A+A) collisions provide a unique
opportunity to study experimentally
different phases of strongly interacting matter and transitions between them,
for the recent review see Ref.~\cite{Florkowski:2010zz}.
In particular, since the discovery of sub-hadronic particles,
quarks and gluons, it was expected that
at high temperature and/or pressure densely packed hadrons will
''dissolve'' into a new phase of
quasi-free quarks and gluons, the quark-gluon-plasma
(QGP)~\mbox{\cite{Ivanenko:1965dg,Itoh:1970uw,Collins:1974ky,Shuryak:1980tp}}.

Years of experimental and theoretical studies of
high energy A+A collisions led to the conclusion
that the QGP exists in nature.
This conclusion is based on a wealth of systematic data on A+A collisions
at very high energies from the CERN Large Hadron Collider
(LHC)~(see, e.g., Ref.~\cite{Braun-Munzinger:2014pya})
and the BNL Relativistic Heavy Ion Collider (RHIC)
(see, e.g., Ref.~\cite{Adams:2005dq}),
and, very importantly
the evidence of the transition
between hadronic matter and QGP (the onset of deconfinement)
at the CERN Super Proton Synchrotron (SPS)
energies~\cite{Afanasiev:2002mx,Alt:2007aa}.

{\bf 2.} The experimental search for the onset of deconfinement
was motivated~\cite{Afanasev:2000dv} by predictions
of the Statistical Model of the
Early Stage (SMES)~\cite{Gazdzicki:1998vd} of A+A collisions.
According to the model the onset of deconfinement in central A+A collisions
should lead to a rapid
change of the energy dependence of several hadron production properties,
all appearing in a common energy domain.
The predicted signals in single hadron properties were
observed~\cite{Afanasiev:2002mx,Alt:2007aa,Gazdzicki:2014sva}.
They indicate that the onset of deconfinement (the beginning
of the mixed phase) is located at $\sqrt{s_{NN}}(OD) \approx 8$~GeV
and the softest point (the end of the mixed phase region) at
$\sqrt{s_{NN}}(SP) \approx 12$~GeV~\cite{Gazdzicki:2014pga},
where $\sqrt{s_\mathrm{NN}}$ denotes collision energy per nucleon pair
in the center of mass system.
However, up to now no convincing signal in fluctuations of event properties
was reported~\cite{Gazdzicki:2015ska}.

Fluctuations are significantly more difficult to study than single hadron
properties. This may explain problems in locating the onset of deconfinement
using fluctuation measurements and it asks for experimental and theoretical
developments in the study of fluctuations in relativistic nucleus-nucleus collisions.

{\bf 3.} This work extends the  model
predictions~\cite{Gazdzicki:2003bb,Gorenstein:2003hk}
on fluctuations of hadron production properties
in relativistic A+A collisions
related to the phase transition.
The predictions have been based on the
SMES model of A+A
collisions~\cite{Gazdzicki:1998vd}.
The extension is needed because new measures of
fluctuations were introduced in the recent years, for reviews see
Refs.~\cite{Gazdzicki:2014sva,Gazdzicki:2015ska}.
In addition, the paper introduces a  more general formalism to
model fluctuations in high energy collisions
which may be helpful in future efforts.

{\bf 4.} Based on the success of statistical and hydrodynamical models
of particle production in high energy collisions
the SMES assumes that the matter created at the early stage
of collisions is in equilibrium.
The created matter  has zero conserved charges, thus, all chemical potentials
are equal to zero, and the fireball energy $E$ and volume $V$ are assumed to vary
from collision to collision according to the probability distribution
function $P(E,V)$.
Consequently, the  energy density,
$\varepsilon =E/V$~,
also may change from collision to collision.
This leads to changes of other properties of matter which,
within the grand canonical ensemble (GCE)
used here, can be calculated using its equation of state (EoS).
In particular,
according to the first and the second
principles of thermodynamics, the entropy change $\delta S$
($\delta$ denotes a deviations from average value) is related
to energy and volume changes as
$T\delta S = \delta E + p \delta V$, which provides
$T\delta S = V \delta \varepsilon + (p + \varepsilon ) \delta V$~,
where $p$ is the pressure. Using the identity $TS = E + pV$ one finds
\begin{equation}\label{eq:entropy}
\frac{\delta S}{S} = \frac{1}{1 +p/\varepsilon} \: \frac{\delta
\varepsilon}{\varepsilon} + \frac{\delta V}{V} \;.
\end{equation}

In case of energy and volume being proportional
and the energy density being constant,  $\delta \varepsilon=0$,
Eq.~(\ref{eq:entropy}) gives:
$\delta S/S=\delta V/V= \delta E/E$.
Thus, the relative fluctuations of entropy are equal to
those of energy and volume, and they are insensitive to the EoS.
In other cases the entropy fluctuations depend on the EoS and thus
on a form of the created matter.

{\bf 5.}
Similarly to entropy, mean multiplicity of particles of a given type
changes with $E$ and $V$. These changes depend on the EoS and
particle properties providing
they lead to changes of $\varepsilon$.
In particular, mean multiplicity of light pions or light
quarks and massless gluons is approximately proportional to entropy
of the produced matter.  When crossing the transition region
the effective number of degrees of freedom increases and thus
the entropy increases faster with increasing $\sqrt{s_{NN}}$.
Consequently,
entropy fluctuations caused by energy density fluctuations will also be modified.
Mean multiplicity of strange  hadrons in the
confined phase or strange quarks in the deconfined phase
is rather sensitive to their masses
and effective
number of degrees of freedom. These quantities
are changed  rapidly when crossing the
transition region by increasing energy density.
This results in a change of fluctuations of mean multiplicity
of strange particles caused by energy density fluctuations.

{\bf 6.}
Within the SMES the fireball created in a high energy collision
is treated as a micro-state which belongs
to the ensemble of all possible micro-states defined by the EoS and
the energy-volume probability density function $P(E,V)$.
In particular,
particle multiplicity is a property of a single event which
varies from event to event even for
fixed values of $E$ and $V$.

In the GCE
these multiplicity fluctuations follow the Poisson distribution.
Four extensive quantities $E$, $V$ and $N$, $N_S$ are considered.
The distribution of the first pair, $P(E,V)$, is assumed, whereas the distribution of
the second pair ${\cal P}(N,N_S)$ follows from fluctuations of
ensemble properties and fluctuations of event properties.

Specially, the scaled variance
of particle multiplicity can be presented \cite{Gorenstein:2003hk} as a sum of the
scaled variance induced by
the initial $E$ and $V$ fluctuations, $\omega_0$, and the scaled variance of particle
multiplicity at fixed $E$ and $V$ values, which is equal to unity
for the assumed Poisson distribution:
\eq{\label{omegaNV}
\omega[N]~=~1~+~\omega_0[N]~,~~~~~~
%
 \omega[N_S]~=~1~+~\omega_0[N_S]~.
}

{\bf 7.}
For the purposes of this work it is sufficient
to characterize the distribution $P(E,V)$ by its five parameters
which include its first and second moments:
\eq{
\langle \varepsilon \rangle~,~~~ \langle V \rangle~,
~~~~~\sdisp{\varepsilon}~,~~~\sdisp{V}~,
~~~~~\frac {\mean{\varepsilon~V}~-~\mean{\varepsilon}~\mean{V}}
{\sqrt{\mean{(\Delta \varepsilon)^2} \mean{(\Delta V)^2}}}~,
}
the latter three, two scaled dispersions and the
correlation coefficient,
are dimensionless.

{\bf 8.}
The SMES formulation and parameter values
(unless otherwise stated) used in the original
papers~\cite{Gazdzicki:2003bb,Gorenstein:2003hk,Gazdzicki:1998vd}
are adopted here.
Thus the ideal gas EoS is used to model
the confined phase, the bag model EoS is used for the QGP phase, and
the first order phase transition between them is assumed.

{\bf 9.}
Since event-by-event 
volume fluctuations cannot be eliminated
in experimental studies of A+A collisions, it is important
to minimize their effect by defining suitable fluctuation measures.
It was shown within the model of independent sources
that one can construct fluctuation measures from the first and
second moments of two
extensive event
quantities, $A$ and $B$, which are
independent of the source number distribution.
%
The measures were referred to as strongly intensive
quantities~\cite{Gorenstein:2011vq}.
The first measure of this type was introduced in
Ref.~\cite{Gazdzicki:1992ri} and then the concept
was generalized~\cite{Gorenstein:2011vq} and extended
to third~\cite{Mrowczynski:1999un} and higher moments~\cite{Sangaline:2015bma}.
Here the predictions of the SMES model will be calculated for
strongly intensive quantities which include the first and second
moments of $A$ and $B$.

{\bf 10.}
Two families of strongly intensive quantities
can be constructed
\cite{Gorenstein:2011vq}:
\eq{\label{eq:delta}
& \Delta[A,B]
 ~=~ \frac{1}{C_{\Delta}} \Big[ \langle B\rangle\,
      \omega[A] ~-~\langle A\rangle\, \omega[B] \Big]~,\\
\label{eq:sigma}
&  \Sigma[A,B]
 ~=~ \frac{1}{C_{\Sigma}}\Big[
      \langle B\rangle\,\omega[A] ~+~\langle A\rangle\, \omega[B] ~-~2\left(
      \langle AB \rangle -\langle A\rangle\langle
      B\rangle\right)\Big]~,
}
where $\omega[X]\equiv (\mean{X^2}-\mean{X}^2)/\mean{X}$.
The normalization factors
$C_{\Delta}$ and $C_{\Sigma}$ are required to be proportional to first moments
of any extensive quantities.
Note that $\Sigma[A,B]$ includes the correlation term $\mean{AB}-\mean{A}\mean{B}$ whereas
$\Delta[A,B]$ does not.

{\bf 11.}
Two selections of the $C_{\Delta}$ and $C_{\Sigma}$ normalization
factors are used in the present paper.
Firstly, the normalization factors equal to
mean of the second argument are assumed:
\eq{\label{eq:norm1}
C_{\Delta}~=~C_{\Sigma}~=~\mean{B}~.
}
%
As pointed out in Ref.~\cite{Gazdzicki:2015ska} this normalization within
the statistical model of the ideal Boltzmann gas in
the GCE (IB-GCE)
formulation
  and $B \sim V$ leads to:
\begin{equation}
   \Delta[A,B]~ = ~\Sigma[A,B]~ =~ \omega^*[A]~,
\label{eq:Bnormalization}
\end{equation}
where $\omega^*[A]$ is the scaled variance of $A$ for a fixed system volume.

Secondly, the normalization
\eq{\label{eq:norm2}
C_{\Delta}~=~\mean{N}~-~\mean{N_S},~~~~~~~~~~~~~ C_{\Sigma}~=~\mean{N}~+~\mean{N_S}
}
will be used for particle multiplicities \cite{Gazdzicki:2013ana}. It
provides $\Delta[N,N_S] = \Sigma[N,N_S] = 1$
in  the IB-GCE.


{\bf 12.}
Within the IB-GCE  and
provided $A$ and $B$ are uncorrelated in a fixed volume,
one finds that
$\omega^*[A]$ and $\omega^*[B]$
can be expressed via $\Sigma[A,B]$ and $\Delta[A,B]$.
Let us introduce the quantities:
\begin{equation}
  \Omega[A,B] ~ \equiv ~ \frac {1} {2}~
    \Big[ \Sigma[A,B] + \Delta[A,B] \Big]~=~\omega[A]-\frac{\cov{A}{B}}{\mean{B}}~,
\label{eq:OmegaA}
\end{equation}
\begin{equation}
  \Omega[B,A] ~ \equiv ~ \frac {1} {2}~
    \Big[ \Sigma[B,A] + \Delta[B,A] \Big]~=~\omega[B]-\frac{\cov{A}{B}}{\mean{A}}~.
\label{eq:OmegaB}
\end{equation}
Here the normalization of
$\Sigma$ and $\Delta$ is given by
Eq.~(\ref{eq:norm1}).
Then one finds \cite{Gorenstein:2011vq, Sangaline:2015bma}
\eq{\label{O}
\omega^*[A]~=~\Omega[A,B]~,~~~~~~~ \omega^*[B]~=~\Omega[B,A]~.
}
%
%
Predictions for $\Omega$ quantities will be also calculated in this paper.

{\bf 13.}
Four extensive event quantities $E$, $V$ and $N$, $N_S$ define
their six pairs:
\begin{equation}
[E,V],
~~~[N,V],
~~~[N_S,V],
~~~[N,E],
~~~[N_S,E],
~~~[N_S,N],
\label{eq:pairs}
\end{equation}
for which the strongly intensive fluctuation measures are calculated.
Within the model the results for the  $[E,V]$ pair depend only
on the assumed distribution $P(E,V)$. The remaining pairs include
at least one extensive event quantity which fluctuations are
dependent on the EoS.

{\bf 14.}
If the energy density $\varepsilon$ remains constant from
event to event, the energy and volume fluctuations are correlated as
$E\sim V$. Within the IB-GCE these fluctuations do not influence
the strongly intensive measures $\Delta$ and $\Sigma$.
The fluctuations of $\varepsilon$ within the SMES
lead, however, to their dependence on $\mean{V}$ and $\sqrt{\mean{(\delta V)^2}}/\mean{V}$.
Namely they are proportional to the term $1+ \mean{(\delta V)^2}/\mean{V}^2$.
For central Pb+Pb collisions one gets $1+ \mean{(\delta V)^2}/\mean{V}^2\cong 1$
and thus in the following calculations the term is set to one.

{\bf 15.}
Based on the SMES model~\cite{Gazdzicki:1998vd} the following
numerical values of the parameters and their
collision energy dependence are assumed.
Mean volume and mean energy density at $\sqrt{s_{NN}} = 10$~GeV
are set to be $\mean{V} = $~350~fm$^3$ and
$\mean{\varepsilon} = $3.2~GeV/fm$^3$, respectively.
Their dependence on collision energy is taken to be
$\mean{V} \sim 1/\sqrt{s_{NN}}$ and
$\mean{\varepsilon} \sim  \sqrt{s_{NN}}~(\sqrt{s_{NN}} - 2~m_N)$.
Scaled dispersion of volume fluctuations is set to $\sdispl{V} = 0$,
which is a good approximation for central Pb+Pb collisions.
Three values of the $\sdispl{\varepsilon}$ parameter are used: 0.17 (solid line),
0.13 (dashed line), and 0 (dotted line).
The value $\sdispl{\varepsilon}=0.17$ is considered
to be the upper limit based on the UrQMD and HSD simulations~\cite{Alt:2007jq}.
The correlation between $E$ and $V$ is set to zero.

\begin{figure}[!htb]
\includegraphics[width=0.49\textwidth]{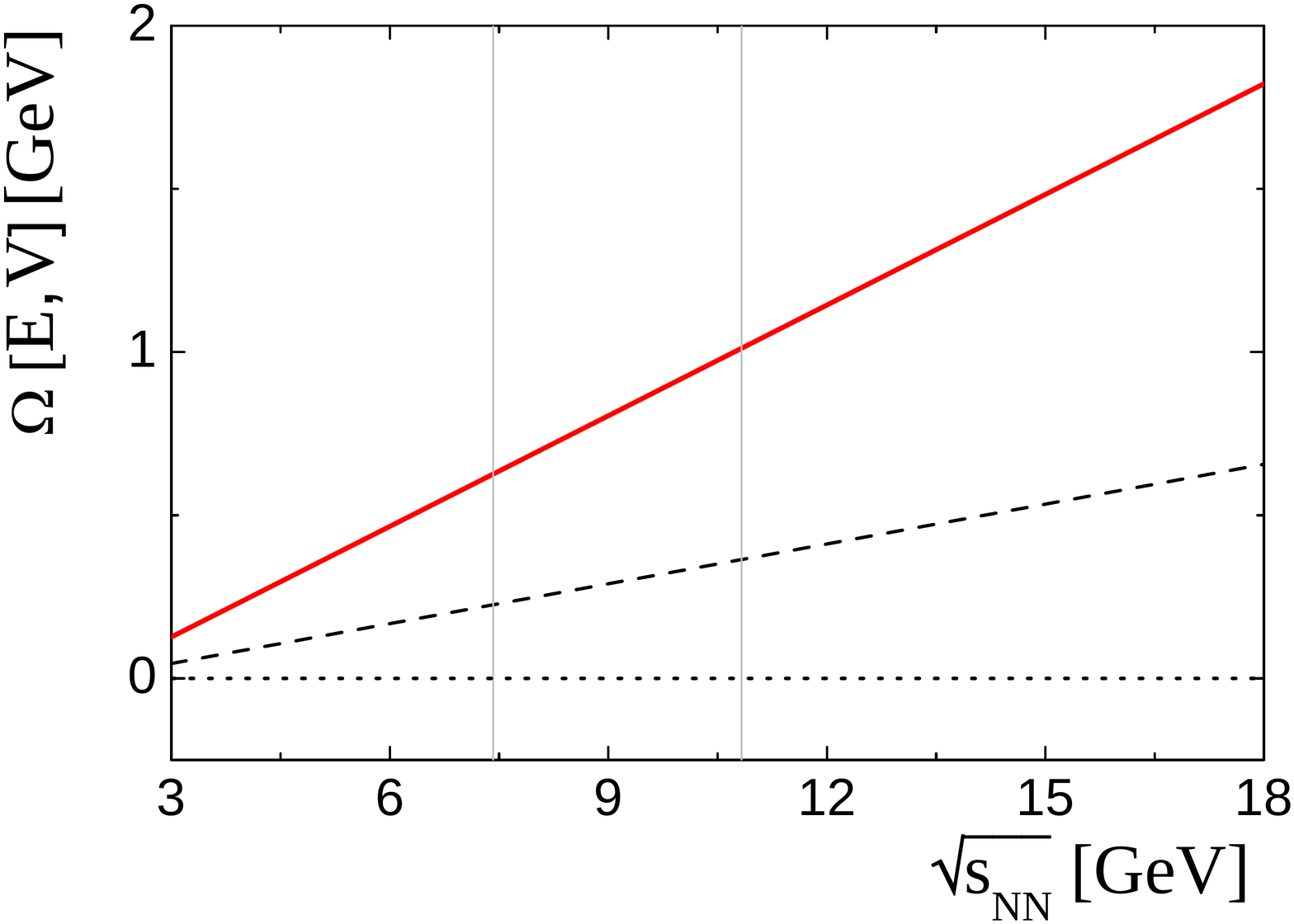}
\caption{$\Omega[E,V]=\Delta[E,V] = \Sigma[E,V] = \omega^*[E]=\mean{(\delta E)^2}/\mean{E}$
as a function of $\sqrt{s_{NN}}$
in central Pb+Pb collisions at the CERN SPS energy range.
See text for the numerical values of the model parameters and their $\sqrt{s_{NN}}$-dependence.
The normalization factors $C_{\Delta} = C_{\Sigma} = \mean{V}$ are used.
The solid, dashed,
and dotted
lines correspond to $\sdispl{\varepsilon}$
equal to 0.17, 0.13,
and 0,
respectively.
%
The vertical lines indicate the beginning (the onset of deconfinement)
and end (the softest point) of the mixed phase region,
$\sqrt{s_{NN}}(OD)  \cong 7.4$~GeV and
$\sqrt{s_{NN}}(SP) \cong 10.8$~GeV, respectively.
}
\label{fig:EV}
\end{figure}

{\bf 16.}
From
Eqs.~(\ref{eq:delta},\ref{eq:sigma},\ref{eq:OmegaA})
one gets
$\Omega[E,V]=\Delta[E,V]=\Sigma[E,V]=\omega^*[E]
= \mean{(\delta E)^2}/\mean{E}$ shown
in Fig.~\ref{fig:EV} as a function of $\sqrt{s_{NN}}$
for the model parameter values
given above.
This result is evidently insensitive to the EoS.

\begin{figure}[!htb]
\includegraphics[width=0.49\textwidth]{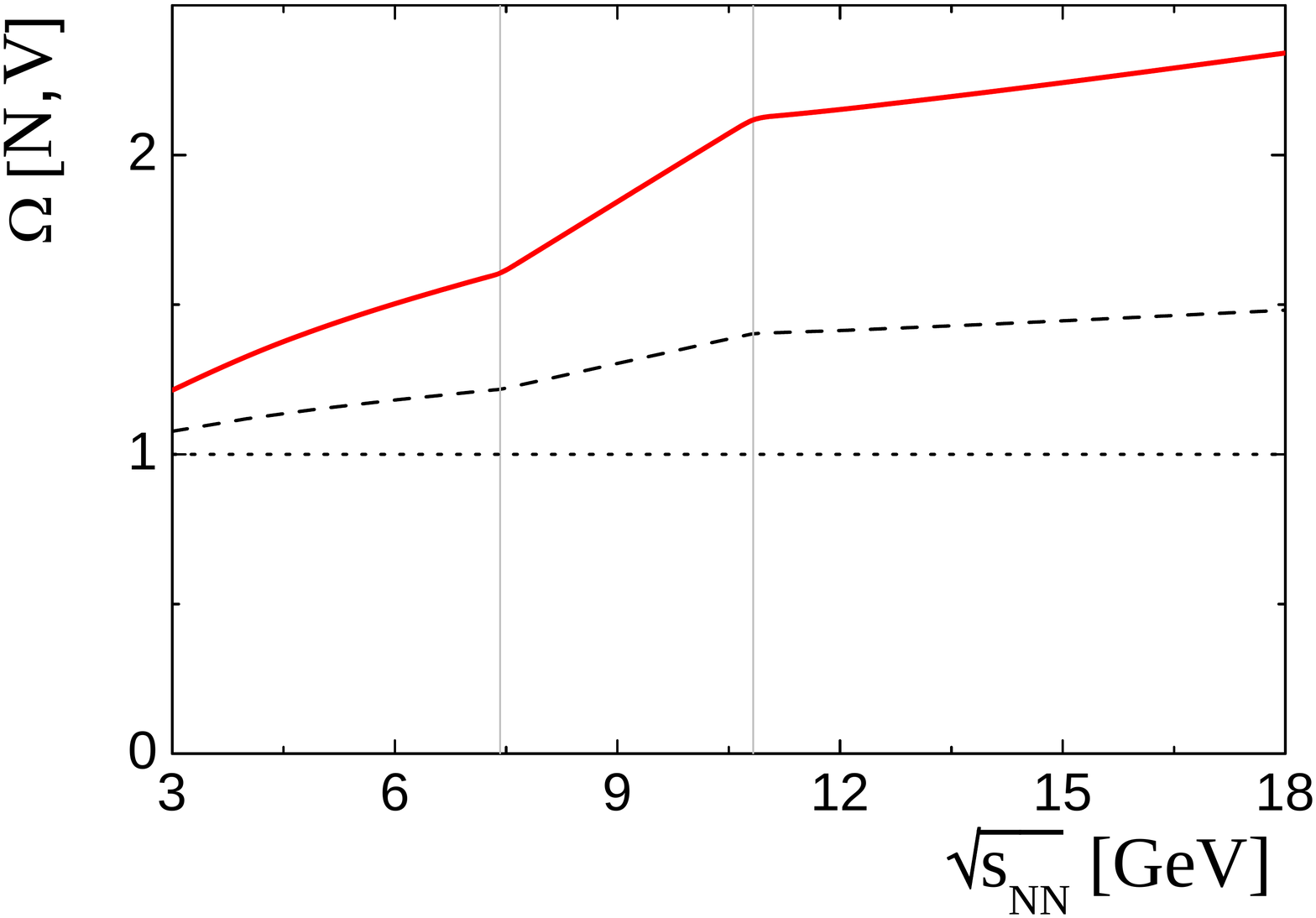}
\includegraphics[width=0.49\textwidth]{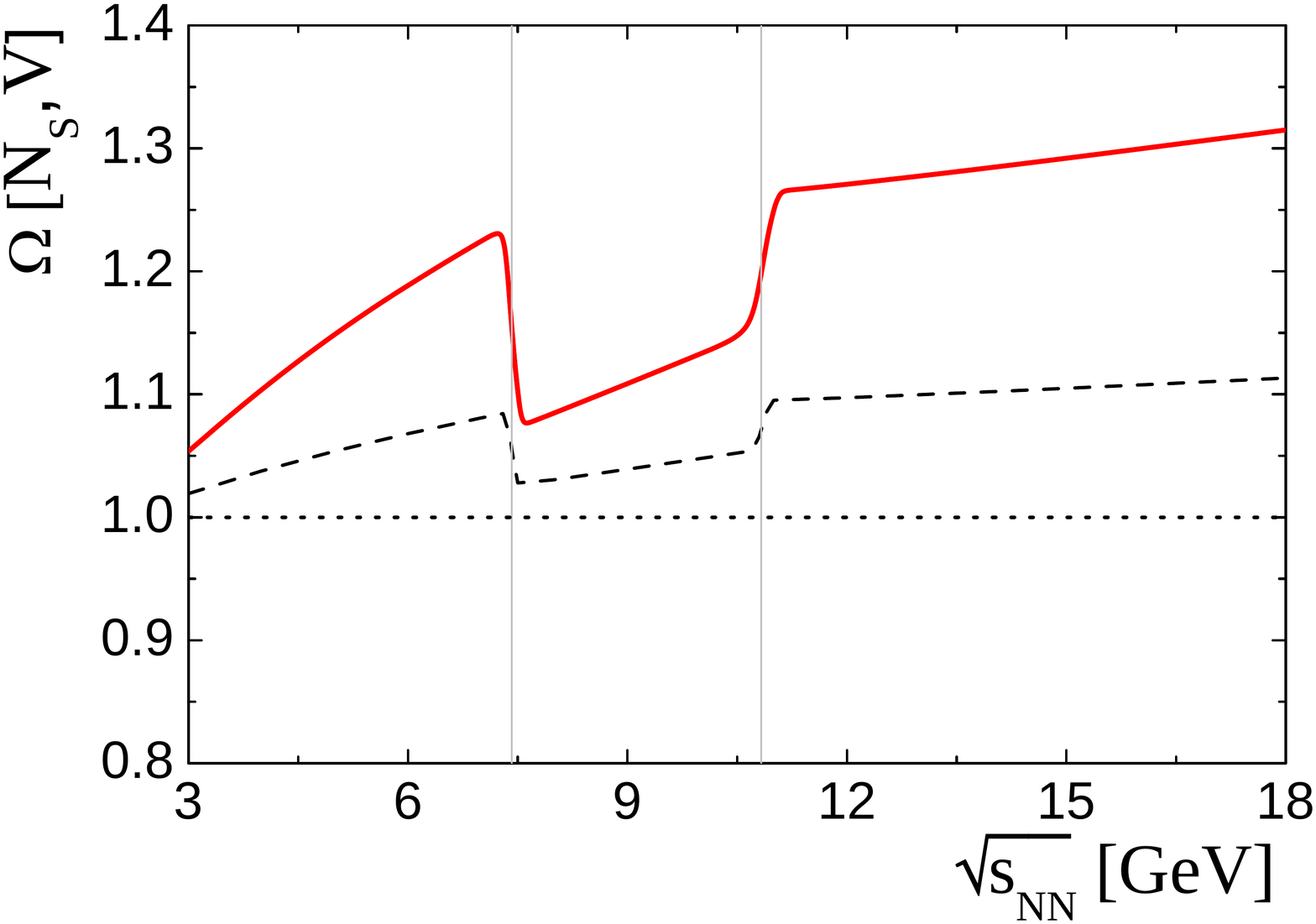}
\caption{The same as in Fig.~\ref{fig:EV} but
for $\Omega[N,V]=\Delta[N,V] = \Sigma[N,V] = \omega^*[N]$ (\itl)
and $\Omega[N_S,V]=\Delta[N_S,V] = \Sigma[N_S,V] = \omega^*[N_s]$ (\itr).
The normalization factors $C_{\Delta} = C_{\Sigma} = \mean{V}$ are used.
%
}
\label{fig:NV}
\end{figure}

\begin{figure}[!htb]
\includegraphics[width=0.49\textwidth]{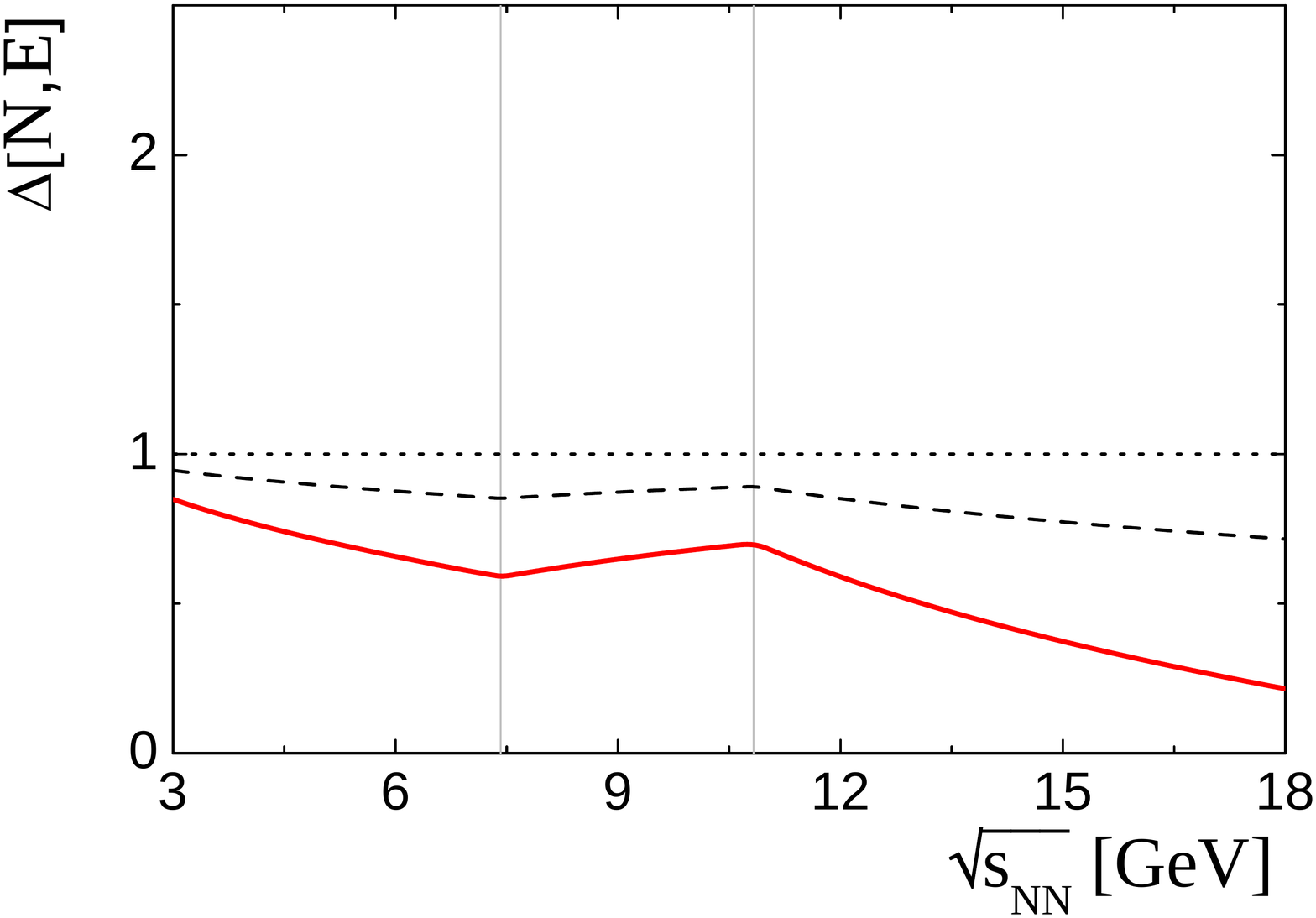}
\includegraphics[width=0.49\textwidth]{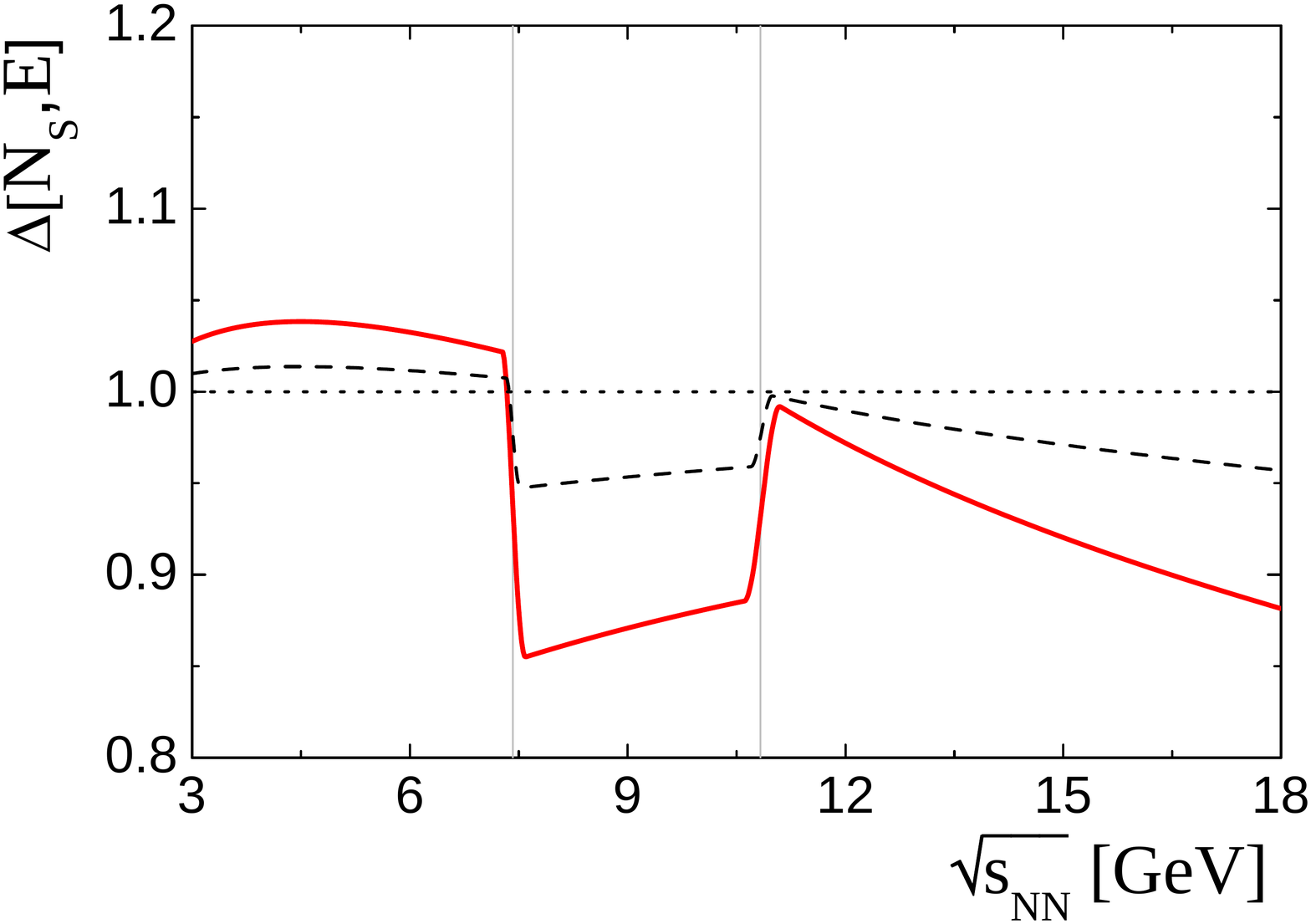}
\includegraphics[width=0.49\textwidth]{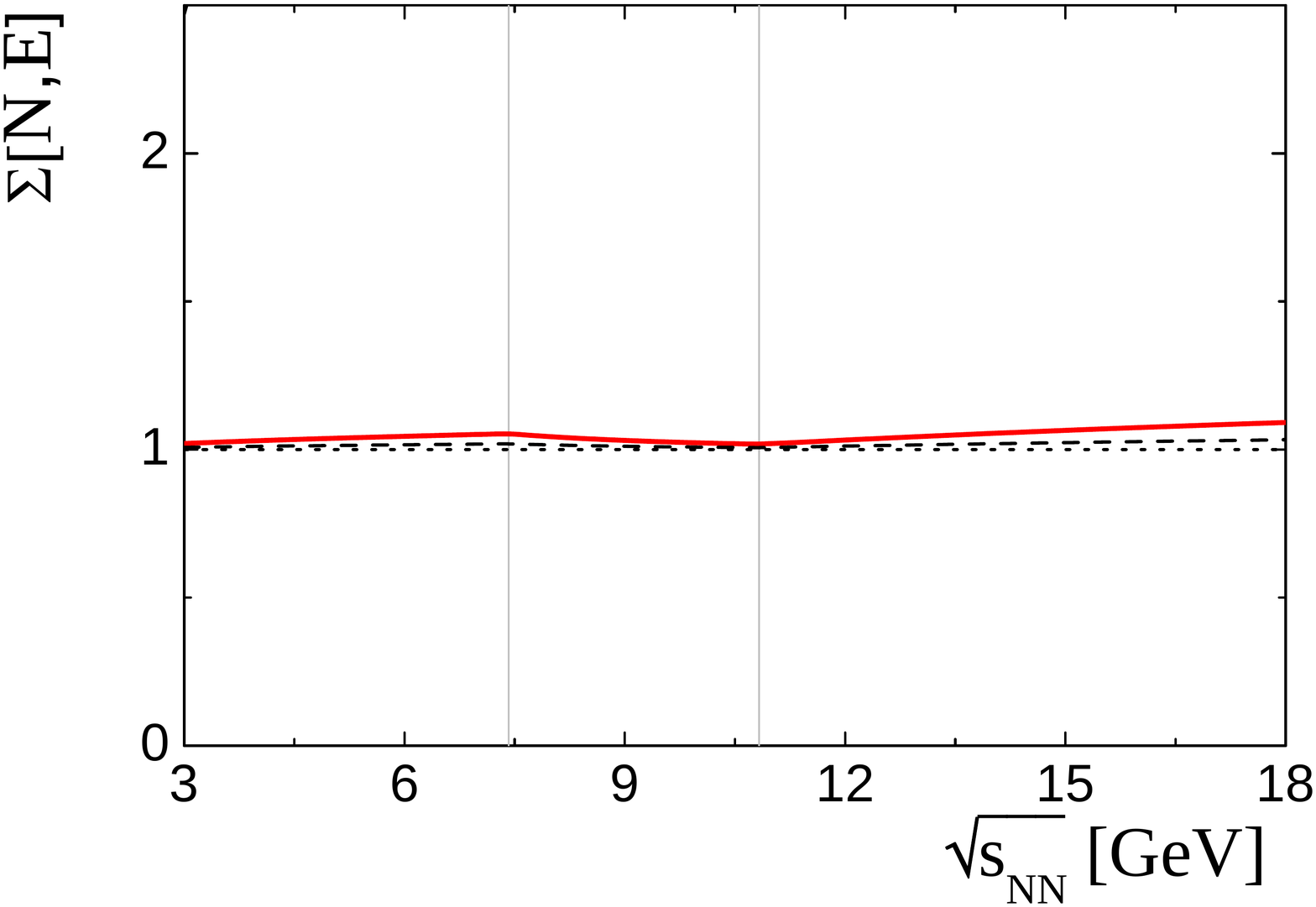}
\includegraphics[width=0.49\textwidth]{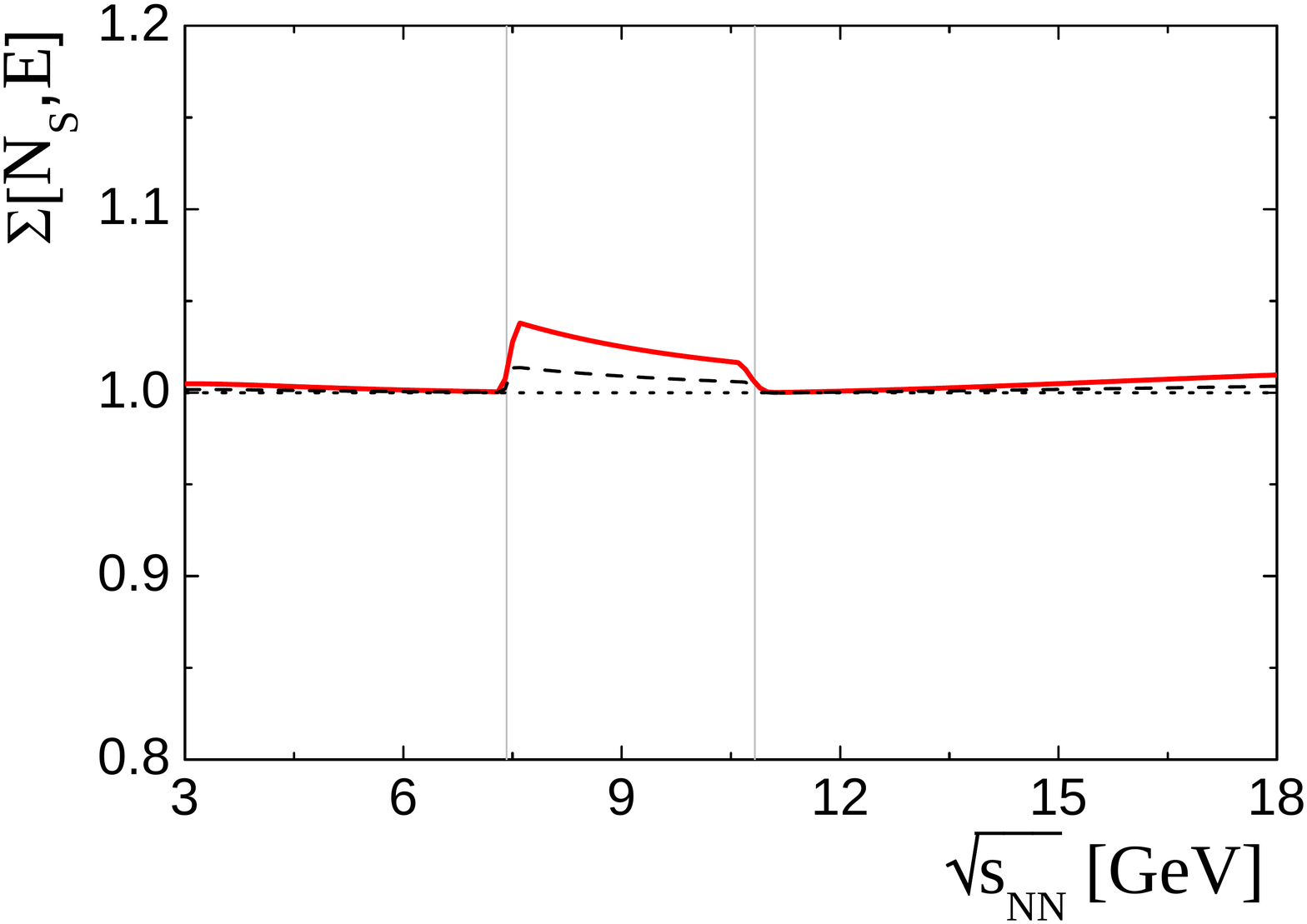}
\caption{The same as in Figs.~\ref{fig:EV} and \ref{fig:NV} but for
$\Delta[N,E]$ and  $\Sigma[N,E]$ ({\it left}),
and
$\Delta[N_S,E]$ and   $\Sigma[N_S,E]$ ({\it right}).
The normalization factors $C_{\Delta} = C_{\Sigma} = \mean{E}$ are used.
%
}
 \label{fig:NE}
\end{figure}

\begin{figure}[!htb]
\includegraphics[width=0.49\textwidth]{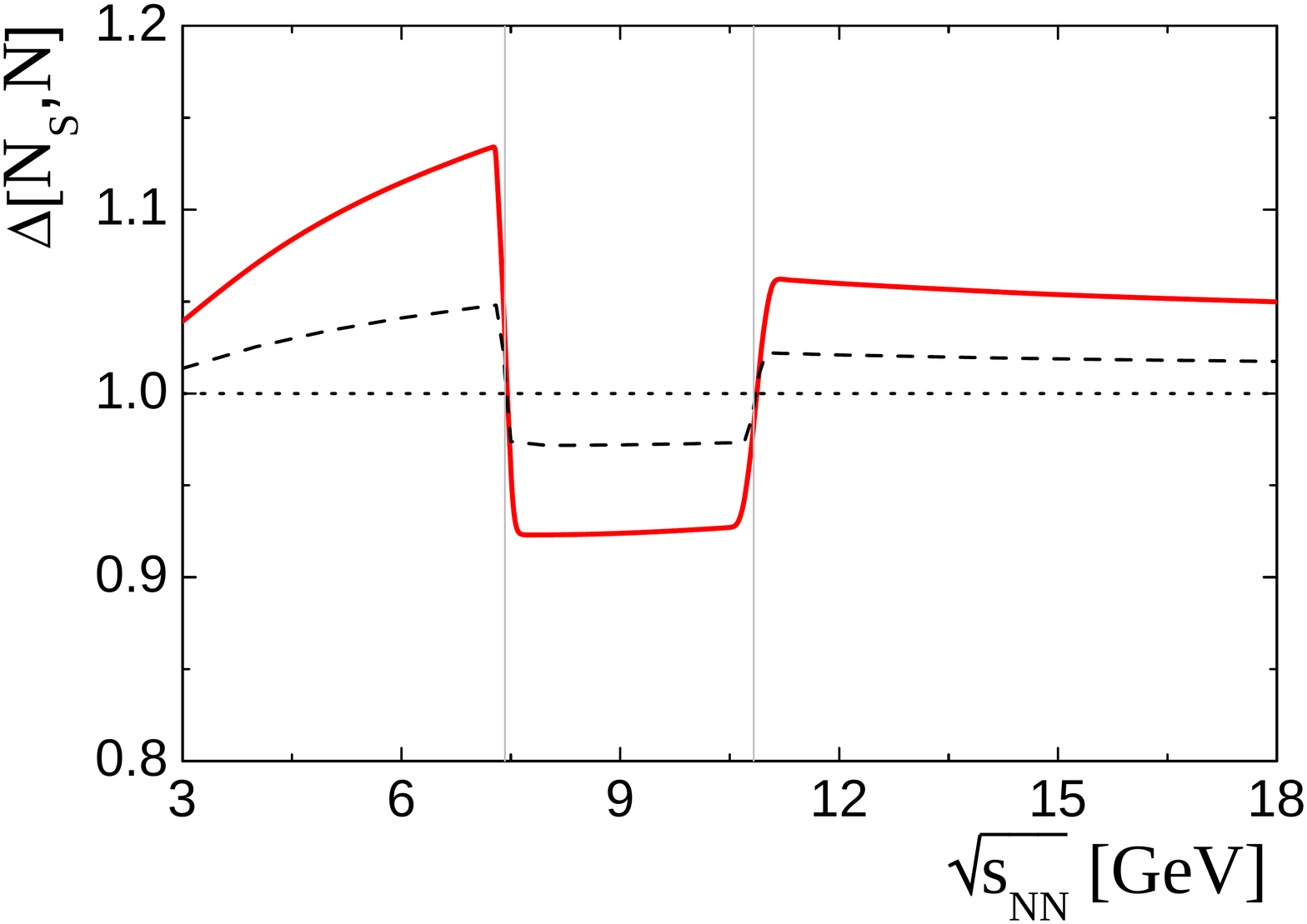}
\includegraphics[width=0.49\textwidth]{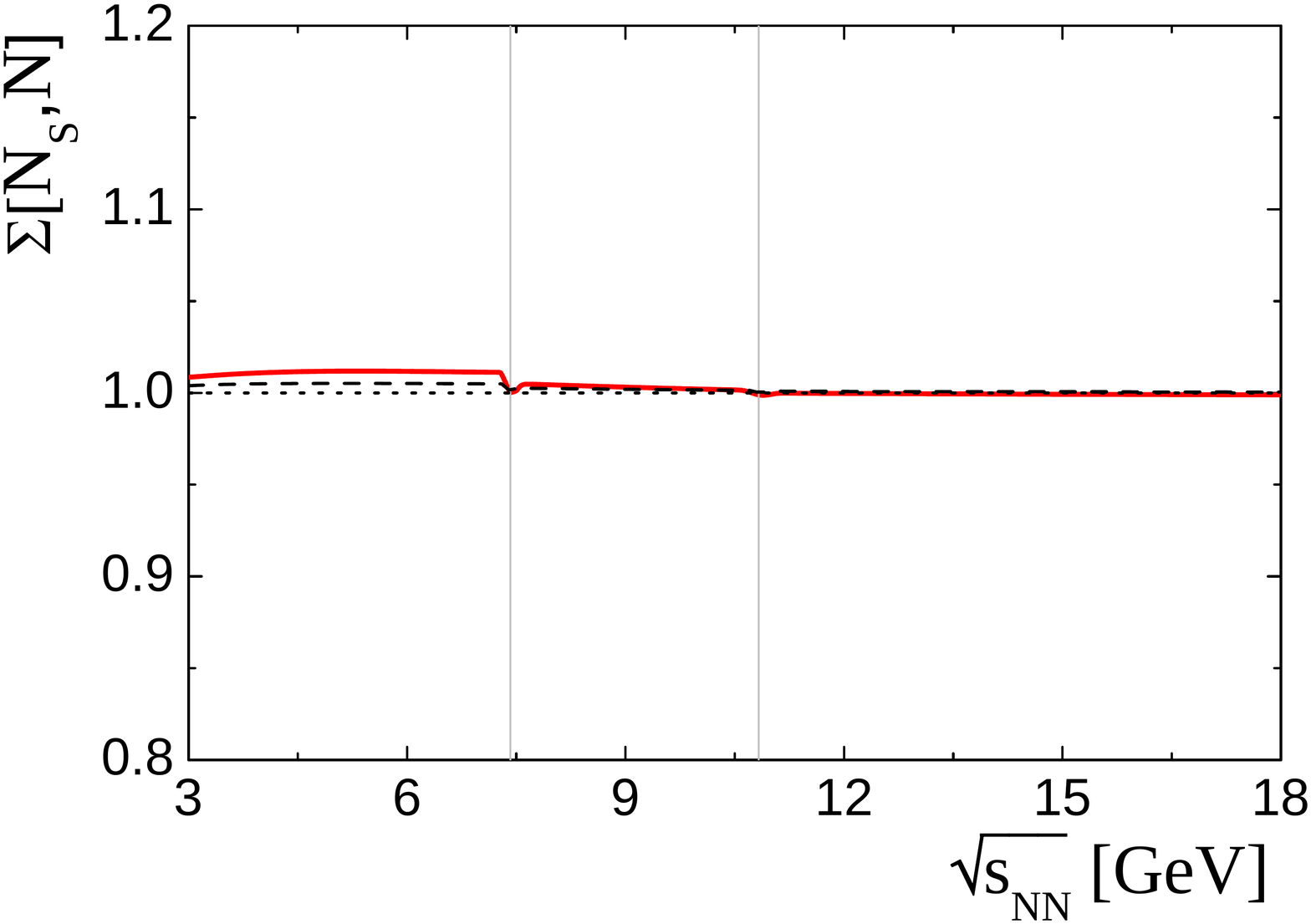}
\caption{The same as in Figs.~\ref{fig:EV}-\ref{fig:NE} but for
$\Delta[N_S,N]$ (\itl) and
$\Sigma[N_S,N]$ (\itr).
The normalization factors $C_{\Delta} = \mean{N} - \mean{N_S}$
and $C_{\Sigma} = \mean{N} + \mean{N_S}$ are used.
%
%
}
 \label{fig:NsN}
\end{figure}

The energy density fluctuations modify
fluctuations of event properties like particle multiplicity and
these modifications are dependent on the EoS and particle type.
On the other hand the EoS and particle properties change
significantly when crossing the
phase transition region. Thus, one expects that the collision
energy dependence of properly selected fluctuation measures
may signal the transition region.
%
The fluctuation measures which are sensitive to the EoS
are plotted in Figs.~{\ref{fig:NV}-\ref{fig:NsN} as  functions of collision energy
in the range which includes the phase transition region.
Note that the $N$ and $N_S$ fluctuations for
$\sdispl{V}~=~0$ and $\sdispl{\varepsilon}~=~0$
are assumed to be Poissonian and uncorrelated.
%

%

{\bf 17.}
The collision energy dependence of
$ \Omega[N,V]=\Delta[N,V] = \Sigma[N,V] =\omega^*[N]$
and $\Omega[N_S,V]=\Delta[N_S,V] = \Sigma[N_S,V] = \omega^*[N_S]$
is shown in Fig.~\ref{fig:NV} for
central Pb+Pb collisions at the CERN SPS energies.
With  $\sdispl{\varepsilon}$ decreasing to zero
the results approach unity as expected in the IB-GCE.
$\Delta$ and $\Sigma$ are equal to each other
for pairs of quantities
$[N,V]$ and $[N_S,V]$. It results from the used approximations:
$N$ and $N_S$ are uncorrelated in the fixed volume,
$\mean{N}\sim V$ and  $\mean{N_S}\sim V$,
and $\mean{(\delta V)^2}/\mean{V}\cong 0$.
%
The overall increasing trend of $\Omega[N,V]$ and $\Omega[N_S,E]$ with $\sqrt{s_{NN}}$
seen in Fig.~\ref{fig:NV} is due to
non-zero values
of $\sdispl{\varepsilon}$.
The modifications of this energy dependence are observed in the region of the
phase transition.
%
However, measurements of system volume (or quantity which
is proportional to volume) are likely
to be experimentally difficult or even impossible.

Figure~\ref{fig:NE} presents
the fluctuation measures $\Delta[N,E]$ and $\Sigma[N,E]$ as well as
$\Delta[N_S,E]$ and  $\Sigma[N_S,E]$
as a function of collision energy.
Similarly to the results presented in Fig.~\ref{fig:NV}
the overall trend caused by the assumed
energy density fluctuations
is modified in the region of the
phase transition.
The most pronounced modification is observed for
$\Delta[N_S,E]$.

Finally, the collision energy dependence of
$\Delta[N_S,N]$
and
$\Sigma[N_S,N]$
is
shown in Fig.~\ref{fig:NsN}.
The normalization (\ref{eq:norm2}) is implied in this case.
The most pronounced modification of the overall trend in the phase transition region
is observed for
$\Delta[N_S,N]$, and a weak one for
$\Sigma[N_S,N]$.

{\bf 18.}
In summary,
predictions on the collision energy dependence of fluctuations
of hadron production properties in central heavy ion collisions in the
range of the phase transition are presented.
They are based on the Statistical Model of the Early Stage and extend
previously published results by predictions for the strongly intensive
quantities.
They are calculated for six pairs of event quantities:
\begin{equation}
[E,V],
~~~[N,V],
~~~[N_S,V],
~~~[N,E],
~~~[N_S,E],
~~~[N_S,N],
\end{equation}
where $E$ and $V$ stand for the system energy and volume, whereas
$N$ and $N_S$ for multiplicities of all and strange particles.
In several considered cases the collision energy dependence is significantly
modified in the phase transition region.
The most pronounced change is seen for $\Delta[N_S,N]$.
This opens a possibility to observe signals of the onset of
deconfinement in data on fluctuations of hadron production
properties.


%
%
\begin{acknowledgments}
We thank Marysia Prior for correcting the paper.
This work was supported by
the Program of Fundamental Research
of the Department of Physics and Astronomy of
National Academy of Sciences of Ukraine (grant ZO-2-1/2015),
the National Science Center of Poland (grant UMO-
2012/04/M/ST2/00816) and the German Research Foundation
(GA 1480/2-2)
\end{acknowledgments}

\bibliography{references}

\newpage

\end{document}